\documentclass[10pt,a4paper]{article}

\usepackage[utf8]{inputenc}
\usepackage[english]{babel}
\usepackage{amsmath}
\usepackage{amsfonts}
\usepackage{amssymb}
\usepackage{graphicx}
\usepackage[left=4cm,right=4cm,top=2cm,bottom=2cm]{geometry}
\usepackage{fourier}

\usepackage{subfig}
\begin{document}

\noindent \begin{Large}\textbf{ Amplitude noise and coherence degradation of
femtosecond supercontinuum generation in
all-normal-dispersion fibers} \end{Large} \\

\noindent \begin{large}Etienne Genier,$^{1,*}$ Patrick Bowen,$^{1}$ Thibaut Sylvestre, $^{2}$ John M. Dudley,$^{2}$ Peter M. Moselund $^{1}$ and Ole Bang$^{1,3}$ \end{large} \\

\noindent \begin{footnotesize}$^{1}$NKT Photonics A/S, Blokken 84, DK-3460, Birker\o d, Denmark\\
$^{2}$Institut FEMTO-ST, UMR 6174 CNRS-Universit\'{e} de Franche-Comt\'{e}, 25030 Besan\c{c}on, France\\
$^{3}$DTU Fotonik, Department of Photonics Engineering, Technical University of Denmark, 2800 Kgs. \\ \hspace{0.15cm} Lyngby, Denmark  \\
$^{*}$ etge@nktphotonics.com \end{footnotesize}\\

\noindent \textbf{Abstract:} Supercontinuum (SC) generation via femtosecond pumping in all-normal dispersion (ANDi) fiber is predicted to offer completely coherent broadening mechanisms, potentially allowing for substantially reduced noise levels in comparison to those obtained when operating in the anomalous dispersion regime. However, previous studies of SC noise typically treat only the quantum noise, typically in the form of one-photon-per-mode noise, and do not consider other technical noise contributions, such as the stability of the pump laser, which become important when the broadening mechanism itself is coherent. In this work, we discuss the influence of amplitude and pulse length noise of the pump laser, both added separately and combined. We show that for a typical mode-locked laser, in which the peak power and pulse duration are anti-correlated, their combined impact on the SC noise is generally smaller than in isolation. This means that the supercontinuum noise is smaller than the noise of the mode-locked pump laser itself, a fact that was recently observed in experiments, but not explained. Our detailed numerical analysis shows that the coherence of ANDi SC generation is considerably reduced on the spectral edges when realistic pump laser noise levels are taken into account.

\section{Introduction}
Highly nonlinear photonic crystal fibers (PCFs) having all-normal-dispersion (ANDi) have only recently emerged as attractive fibers to generate low-noise, octave-spanning supercontinua (SC) [1-4]. This is due to the fact that these PCFs are difficult to fabricate (e.g, requiring submicron holes), and that broadband low-noise ANDi SC generation requires high peak power femtosecond (fs) lasers. It has been shown that low-noise ANDi SC generation cannot be achieved using long pulses because the Raman effect is just as noisy as modulation instability \cite{Mol1}. Despite these obstacles, fs-pumped ANDi SC generation has received significant attention because of its ability to generate temporally coherent SC pulses; a feature that is unachievable in the anomalous dispersion regime. This gives such systems potential in a range of fields including optical coherence tomography (OCT), optical metrology, photo-acoustic imaging, and spectroscopy [6-11].  \\

The reason for this high coherence comes from the coherent spectral broadening mechanisms of self-phase modulation (SPM) and optical wave-breaking (OWB) \cite{Heidt2, Finot}. In fs-pumped ANDi SC generation these mechanisms are considerably more efficient than the incoherent nonlinear effects of noise-seeded modulation instability and stimulated Raman scattering \cite{Heidt1}, allowing for octave spanning SC generation in which these incoherent effects are suppressed for sufficiently short fiber lengths and input pulse durations \cite{Heidt2, Ivan}.\\

Several theoretical studies about SC coherence have been reported in anomalous dispersion fiber, covering pulse durations from fs up to continuous wave, and including several kinds of noise, such as one-photon-per-mode (OPM) quantum noise added as pure phase noise in the frequency domain, Wigner-representation based quantum noise added in the time domain as amplitude and phase noise, Raman noise, and polarization noise [15-26]. This is not the case for theoretical studies of ANDi SC generation, which have mostly considered only the OPM noise \cite{Heidt1, Heidt2, Dudley1}, except for a recent work of Gonzalo \emph{et al.} \cite{Ivan}, in which relative intensity noise (RIN) was experimentally and numerically compared, for ANDi SC generation with 170 fs pump pulses. In this study, it was found that OPM noise was too weak to describe the experimentally observed noise, whereas adding 1 \% amplitude fluctuations of the laser gave better agreement with the experimental results. This important result underlines that while pump laser noise traditionally has contributed little to the noise levels in incoherently broadened anomalous dispersion fiber SC generation \cite{Dudley1, Mol}, it is extremely important in coherently broadened ANDi SC generation. \\

Unfortunately Gonzalo \emph{et al.} just briefly mentioned the effect of pure amplitude noise of the pump laser for one specific pulse duration and fiber length and no specific simulation including the amplitude noise was shown and no general study conducted \cite{Ivan}.
In this work, we therefore present a comprehensive study of the influence of the noise of a mode-locked pump laser on the coherence and RIN of the SC spectrum generated in an ANDi fiber. We focus on single polarization SC generation because it allows us to more clearly demonstrate the impact of technical laser noise. In particular we consider both fluctuations in the amplitude and pulse length, which typically are anti-correlated in a mode-locked laser.  We show both numerically and analytically that combined their impact on the SC noise is generally smaller than in isolation, which explains that the SC noise generally is smaller than the noise of the pump laser itself, as was observed recently \cite{Shreesha}. \\

These results show that the limits for high coherence, in terms of pulse duration and fiber length, suggested in previous publications, change substantially when technical pump laser noise is included. Indeed, we find that while a high coherence can be maintained for a pulse duration below 1.2 ps in a system without technical pump laser noise and only with one polarization \cite{Heidt2}, a pulse duration of $\sim$ 50 fs will begin to lose coherence even when a relatively low noise pump laser is properly described.  

\section{Numerical model and noise sources}

In the numerical study we use a single-polarization scalar model in the form of the standard generalized nonlinear Schr\"odinger equation to model the propagation of the envelope function $A=A\left(z,T\right)$, with initial condition $A\left(0,t\right)=\sqrt{P_{0}}{\rm sech}\left(t/T_{0}\right)$, in a highly nonlinear single mode optical fiber. This includes dispersion (described by a Taylor expansion up to $\beta_{10}$), spectrally dependent linear loss $\alpha(\omega)$ and the nonlinear response of the material as well as the dispersion of the nonlinearity and the Raman response, $R\left(T\right)$ \cite{Dudley1}:
\begin{equation}
\frac{\partial A}{\partial z} = -\frac{\alpha(\omega)}{2}A + \sum_{k\geq 2} \frac{i^{k+1}}{k!} \beta_{k} \frac{\partial^{k} A}{\partial T^{k}} + i\gamma \left( 1+ i\tau_{0} \frac{\partial}{\partial T} \right) \left( A \int_{-\infty} ^{+\infty} R(T') |A(z,T-T')|^{2}dT' \right),
\end{equation}
where $\gamma$ is the nonlinear coefficient and $\tau_{0}=1/\omega_{0}$ represents the characteristic time scale of self-steepening \cite{Dudley1}. The Raman response is modelled using the real experimentally measured Raman gain profile as described in \cite{bra}. \\

In this study we consider two sources of noise. The first is the well-known quantum noise $\delta_{\textrm{QN}}$, modelled semi-classically as the standard OPM noise, which is added to the initial condition in the Fourier domain as one photon of energy $\hslash\omega_m$ and random phase $\Phi_m$ in each spectral bin with angular frequency $\omega_m$ and bin size $\Delta\Omega$ \cite{smith}. The OPM noise is in the frequency domain given by 
$\delta_{\textrm{QN}}=\sqrt{h\omega_{m}/\Delta\Omega } \,\textrm{exp}(i2\pi\Phi_m) $, where $h$ is Planck's constant and $\Phi_m$ is a random number Gaussian distributed in the interval [0,1]. In our simulations we used $2^{15}$ points and thus m$\in$$[1,2^{15}]$.\\

The second noise source is the laser amplitude fluctuations described by $\delta_{\textrm{AN}} = \Psi$, where $\Psi$ is a single random value for each input pulse, extracted from a Gaussian distribution with a unit mean and a standard deviation equal to the rms amplitude noise of the modelled laser, given by the manufacturer of the laser. In this paper, we will consider a range of rms amplitude noises going from 0.1\% to 2\% (0.2\% representing a Onefive Origami 10 fs laser and 1\% a Fianium fs laser, both from NKT Photonics).  \\ 

To correctly consider the effects of laser amplitude fluctuations on the SC, it is important to take into account any correlated fluctuations of the pulse duration, which occur in a mode-locked laser. This is important to model, as pulse duration fluctuations will subtly affect the efficiency of SPM, and thus the output spectral shape and width. To estimate the correlation of the fluctuations, we use the deterministic correlation between average power and spectral bandwidth experimentally measured in an Origami 10 fs mode-locked laser. Assuming a fixed repetition rate and a state of anomalous dispersion soliton mode-locking, producing approximately transform-limited sech-shaped pulses, we then find the linearized relation between peak power and pulse length. Assuming then that this relation holds during the fluctuations we obtain the following relationship:
\begin{equation}
\delta _{T_{0}}=-0.8 *\left(\delta_{\textrm{AN}} - 1 \right),
\end{equation}
where $\delta_{\textrm{T}_{0}}$ is a Gaussian distribution centered in 0 having a standard deviation equal to 0.8 times $\delta_{\textrm{AN}}$. The peak power and pulse duration of a mode-locked laser are thus anti-correlated, which is extremely important for the SC noise, ass we shall see in the following. We note that the value of 0.8 is specific to the laser considered, but can be generalized for other kinds of lasers.
With both noise terms included our initial condition becomes: 
\begin{equation}
A\left(0,t\right)=\sqrt{P_{0}} \delta_{\textrm{AN}} \textrm{sech} \left(t /(T_{0}(1 -0.8(\delta_{\textrm{AN}}-1) ))\right) + F^{-1}\{ \delta_{\textrm{QN}} \},
\end{equation}
\noindent where $F^{-1}$ represents the inverse Fourier transform. Here and in the following the Fourier transform of a function is denoted by a tilde and defined as \\ $\tilde{A}(z,\omega)=\int_{-\infty}^\infty A(z,t)\exp(i[\omega-\omega_0]t)dt$, where $\omega_0$ is the pump angular frequency.   

\section{Results}

As explained above, we want to focus on just the effect of laser technical noise compared to the conventional OPM noise and therefore use a single-polarization scalar code. This is still highly accurate for polarization maintaining (PM) fibers, which is why we choose the specific PM ANDi PCF NL-1050-NE-PM from NKT Photonics. This PCF has a relative hole size of $d/L=0.45$, a small hole-to-hole pitch of 1.44 $\mu$m, and a stress-induced birefringence of $4\times10^{-4}$. Its classical ANDi dispersion profile, shown in Fig 1(a), was calculated with COMSOL and confirmed experimentally in the region 900-1300 nm using white light interferometry. The measurements of dispersion were done without controlling polarization, which is typically sufficient for stress-induced birefringence that does not significantly alter the mode profile. As expected, the small holes of the PCF give a confinement loss edge significantly below the material loss edge, here found to be at 1450 nm using COMSOL (see Fig 1(a)). This will significantly influence the long wavelength part of the intensity and noise profiles, as we will see in the following. The dispersion has a maximum of -13ps/nm/km at 1040~nm and is rather symmetrical within the low-loss window. However, we pump just above the maximum dispersion at 1054~nm, because we want to consider realistic noise values of a specific laser -- the Origami 10. \\ 

\begin{figure}[t]
   \begin{minipage}[c]{.24\linewidth}
   \includegraphics[scale=0.56]{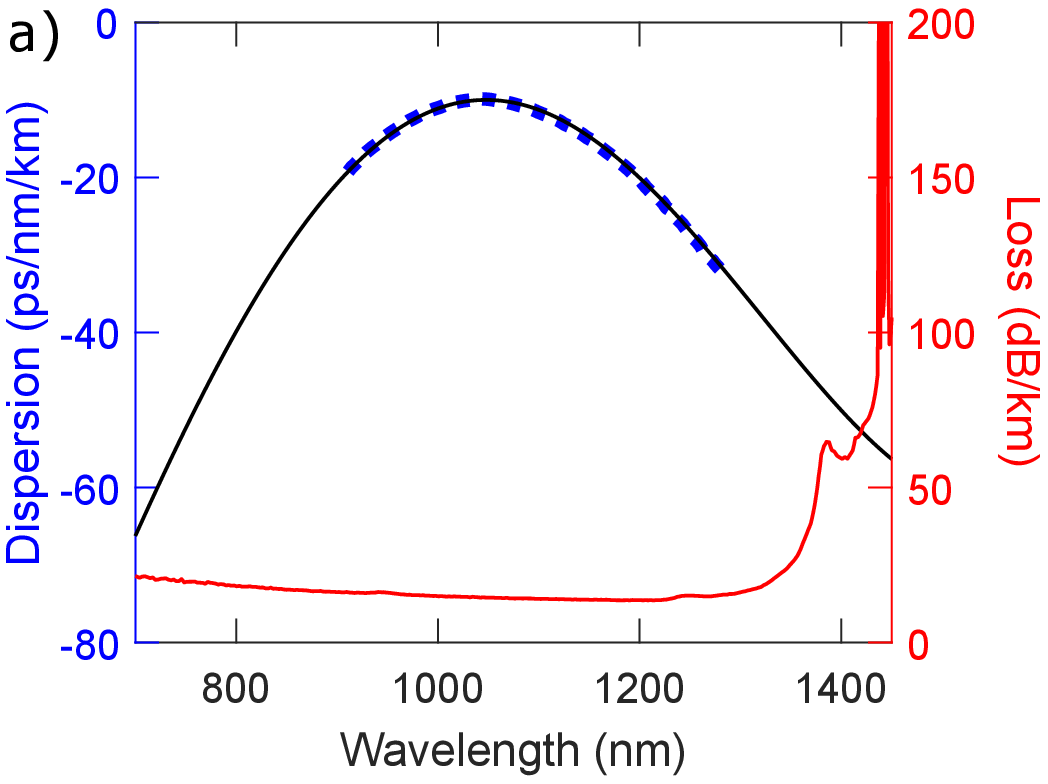} 
   \end{minipage} \hfill
   \begin{minipage}[c]{.45\linewidth}
      \includegraphics[scale=0.57]{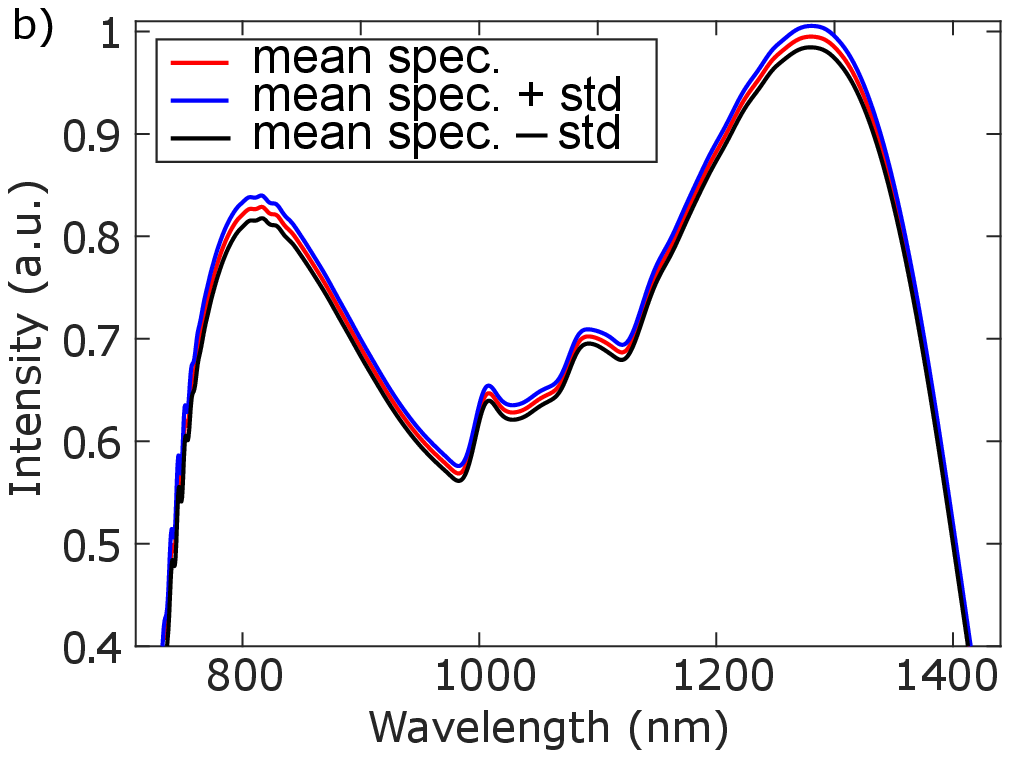}
   \end{minipage} 
      \caption*{Fig 1. (a) Measured (dashed blue) and modelled (solid black) dispersion profile, and fiber losses (solid red) of the NL-1050-NE-PM ANDi PCF. (b) Numerical SC spectra generated in 1 m of ANDi fiber with a 1054 nm pump with an average peak power and pulse duration of $P_0$=100kW and $T_0$=50 fs, respectively. An amplitude noise of 0.5\%  was used, corresponding to a pulse duration noise of 0.4\%. \vspace{-1\baselineskip}}
\end{figure} 

In Fig. 1(b) we show the numerically found SC spectrum out of 1m of ANDi fiber generated by a pump laser with an average peak power and pulse duration of $P_0$=100kW and $T_0$=50 fs, respectively, and an amplitude and corresponding anti-correlated pulse duration noise of 0.5\% and 0.4\%, respectively. We used an ensemble of 20 simulated pulses to calculate the mean and standard deviation for each wavelength and show in Fig. 1(b) the mean and the mean +/- the standard deviation. The results show that with only a small pump laser amplitude noise of 0.5\%, fluctuations in the SC spectrum are already noticeable. The calculations were repeated with 40 and 80 pulses in the ensemble and no noticeable change was found, which means that the statistics can be trusted.\\

In ANDi SC generation, it has been shown that the pulse duration, fiber length, and peak power have a critical influence on the noise properties \cite{Heidt2, Ivan}, e.g., the anticipated coherent spectra are only obtained for sufficiently short pulse durations and fiber lengths. To see the effect of pump laser amplitude noise on the coherence and the requirements on the fiber length and pulse parameters, we simulate a wide parameter space and calculate for each case the spectrally averaged coherence given by \cite{Heidt2, Dudley1}:
\begin{equation}
|g_{12}(\omega)| = \left | \frac{\left \langle   \tilde{A}_i^{*}(\omega) \tilde{A}_j(\omega)  \right \rangle_{i\neq j}} { \sqrt{\left \langle \left | \tilde{A}_i(\omega)  \right |^{2} \right \rangle \left \langle \left | \tilde{A}_j(\omega)  \right |^{2}  \right \rangle  }  } \right |, \quad \left \langle \left | g_{12} \right | \right \rangle = \frac{\int_{0}^{\infty	}\left |   g_{12}(\omega) \right | \left \langle \left | \tilde{A}_i(\omega) \right |^{2} \right \rangle d\omega }{\int_{0}^{\infty} \left \langle \left | \tilde{A}_i(\omega) \right |^{2} \right \rangle d\omega  }
\end{equation}
where $\tilde{A}_i(\omega)=\tilde{A}_i(z,\omega)$, $\langle\tilde{A}_i(\omega) \rangle$ denotes an ensemble average, and subscripts $(i,j)\in[1,20]$ runs over the ensemble. This gives a single number characterizing the coherence for each set of parameters. In Fig. 2(a) we plot the colour coded spectrally averaged coherence $\langle \left| g_{12} \right | \rangle$ versus the pulse duration and fiber length, for a fixed weak amplitude noise of 0.3\% (giving an anti-correlated pulse duration noise of 0.24\%). As is typical of ANDi-SC sources, the coherence decreases when either the pulse duration or fiber length is increased \cite{Heidt2, Ivan}. However, unlike earlier single-polarization studies without amplitude noise, we observe in Fig 2(a) a considerably limited range of parameters where the noise is low, defined as when the spectrally averaged coherence $\langle \left | g_{12} \right | \rangle$ is higher than 0.9 (dashed line in Fig. 2(a)). In particular, good coherence is seen to require pulse durations below 100 fs, which is an order of magnitude shorter than the corresponding limit found without this even weak laser amplitude noise of 0.3\% \cite{Heidt2, Ivan}.\\ 

In Fig. 2(b) we show the low noise limit $\langle \left | g_{12} \right | \rangle = 0.9$ for amplitude noise levels between 0.1-1~\% (pulse duration noise from 0.08 to 0.8\%). From Fig. 2(b) we see that for amplitude noise levels higher than 1\% an average coherence of 0.9 cannot be obtained for any reasonable fiber length. From this, we can see clearly the dramatically limiting effect that the addition of standard laser noise levels has on the coherence parameter space. Significantly, Fig. 2(b) does not show a contour line for the case when only OPM noise is present because there is no loss of coherence observable until pulse durations as long as 1.2 ps, as also shown by Heidt \emph{et al.} \cite{Heidt2}.\\

\begin{figure}[t]
   \begin{minipage}[c]{.46\linewidth}
   \includegraphics[scale=0.62]{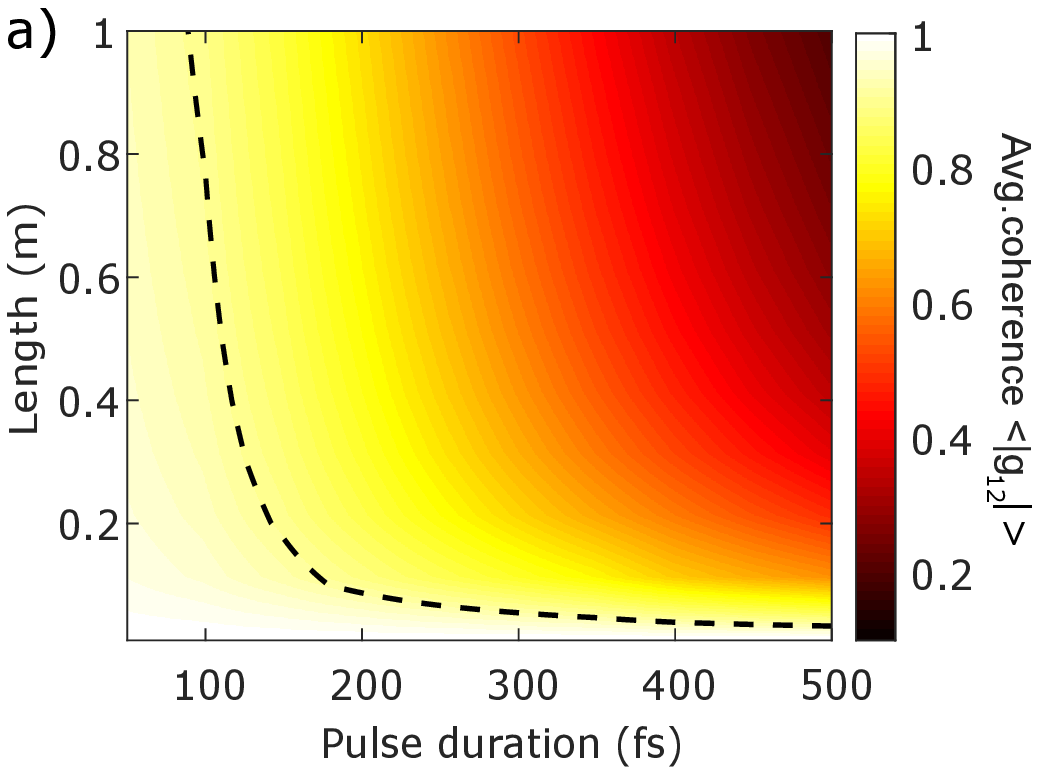} 
   \end{minipage} \hfill
   \begin{minipage}[c]{.46\linewidth}
      \includegraphics[scale=0.59]{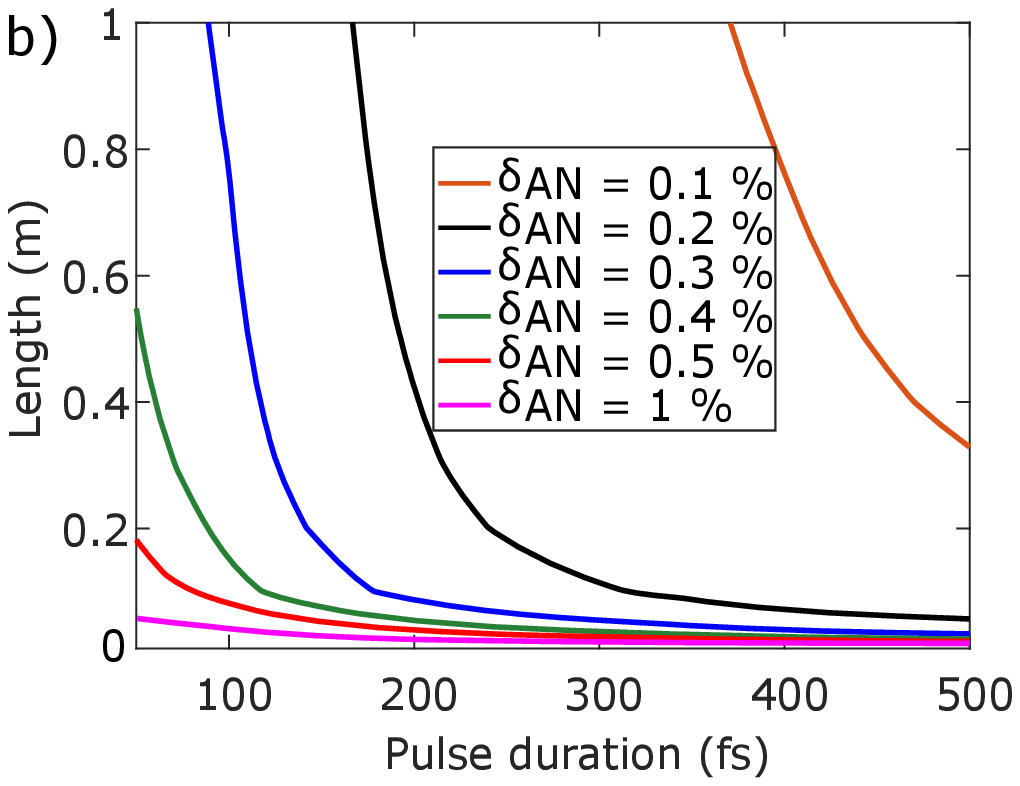}
   \end{minipage} 
\caption*{Fig 2. (a) Average spectral coherence $\langle \left | g_{12} \right | \rangle$ of SC pulses generated with $P_0$=100 kW peak power pump pulses as a function of pump pulse duration $T_0$ and propagation distance for an amplitude noise value of 0.3 \% (pulse duration noise 0.24\%). The dotted line indicates the limit $\langle \left | g_{12} \right | \rangle$=0.9. (b) Limit $\langle \left | g_{12} \right | \rangle$=0.9 for a range of amplitude noise values from 0.1-1 \% (pulse duration noise 0.08-0.8\%). \vspace{-1\baselineskip} }
\end{figure}

It is interesting to look into the specific spectral structure of the ANDi noise. To do so we consider the experimentally more relevant RIN, which is typically used to characterize the noise of an SC source. The frequency-dependent profile, RIN($\omega$), is defined as \cite{Ivan}: 
\begin{equation}
{\rm RIN}(\omega) =  \sqrt{\left \langle \left ( \left |   \tilde{A}(\omega) \right |^{2} - \left \langle   \left |   \tilde{A}(\omega) \right |^{2} \right \rangle \right )^{2} \right \rangle }/ \left \langle   \left |   \tilde{A}(\omega) \right |^{2} \right \rangle.  
\end{equation}
Figure 3(a) shows the mean of the intensity spectrum and the RIN($\omega$) of an SC generated with 100kW, 50fs pulses at 1054nm for a reasonably long fiber length of 1 m for pump laser amplitude noise levels of 0.1, 0.5, and 1.0\%. We again used 20 pulses in the ensemble. \\

We see that in all cases the RIN is low for the majority of the bandwidth, but increases strongly at the edges as expected. Two things are interesting to note:
First, for the weaker amplitude noise levels of 0.1 and 0.5\%, the RIN at the red edge is significantly lower than the RIN at the corresponding blue edge (e.g., the 0.6 intensity level for 0.5\% noise). This is because the loss at the red edge is much higher than at the blue edge due to the strongly increasing confinement loss at 1450 nm. Indeed, if we omit fiber losses from our model, we find similar RIN values at the two spectral edges. We note that the impact of the long-wavelength loss edge was studied previously in conventional anomalous dispersion SC generation, in which it was also found to reduce the noise, here by suppressing rogue wave generation \cite{daniel}. \\

Secondly we see in Fig. 3(a) a signature of peaks in the spectrum being correlated with peaks in the RIN profile. This correlation becomes even more pronounced at shorter fiber lengths where SPM is dominating, and generates a periodic spectrum and RIN profile, as is visible in the initial 20 cm of the propagation shown in Fig. 3(b), which is a colormap of the spectral RIN evolution as a function of the fiber length. We note that while the average RIN increases during the propagation, the spectrally resolved RIN in the central region of the spectrum decreases with the propagation, with the higher RIN quotients being pushed to the edges of the spectrum. \\

In Fig. 4(a) we show the mean spectrum and the RIN profile after 10 cm of propagation and draw vertical lines from the peaks in the RIN profile to the corresponding point in the mean spectrum (OPM noise was removed for better clarity). The correlation is very clear here as a closely matched periodicity and it appears that when SPM is dominating the influence of amplitude and pulse duration noise is to generate strong RIN peaks close to the fringes with maximum slope in the spectrum. We again used 20 pulses in the ensemble and the calculations were repeated with 40 and 80 pulses in the ensemble, which revealed no noticeable change. This means that the RIN statistics is converged and can be trusted. \\

\begin{figure}[t]
   \begin{minipage}[c]{.2\linewidth}
   \includegraphics[scale=0.64]{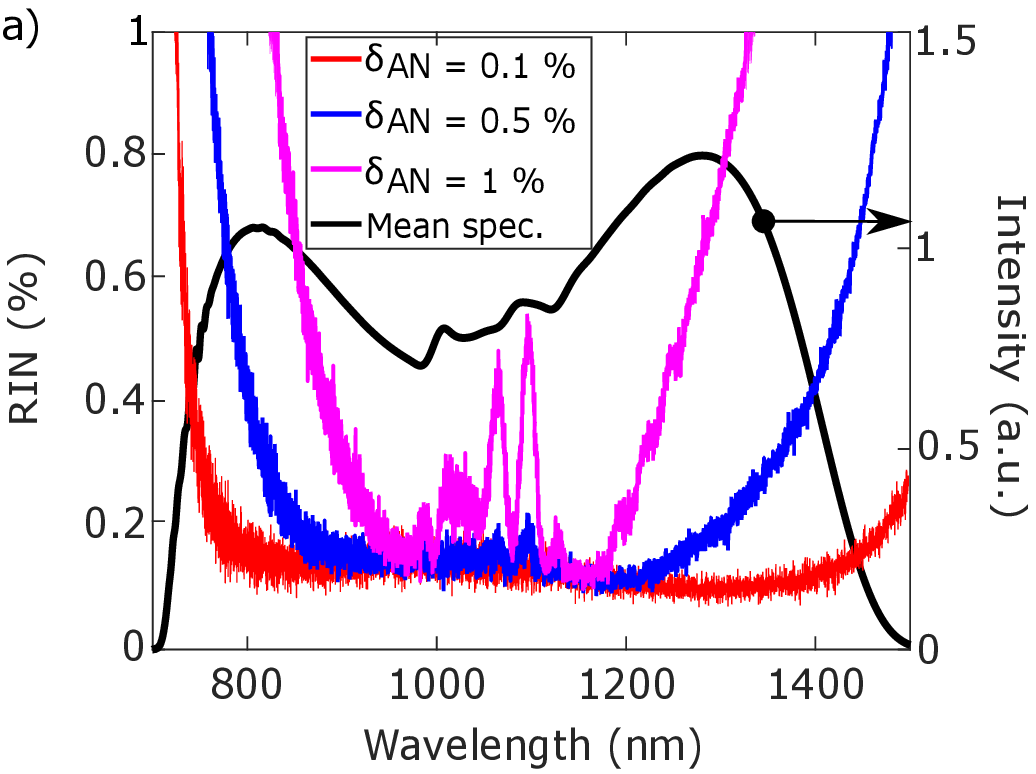} 
   \end{minipage} \hfill
   \begin{minipage}[c]{.48\linewidth}
\includegraphics[scale=0.64]{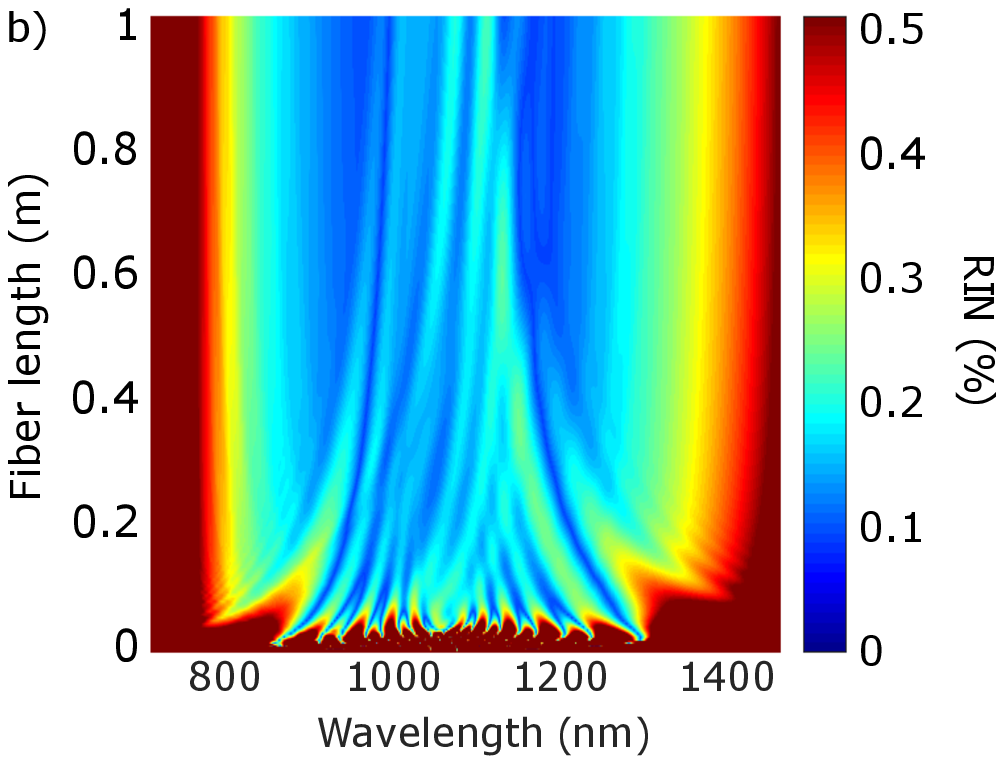} 
   \end{minipage} 
\caption*{Fig 3. (a) RIN profiles  for different amplitude noise values and mean spectral profile out of 1 m of ANDi fiber pumped with 100 kW peak power, 50 fs long pulses at 1054 nm. (b) Evolution of the RIN along the fiber length for an amplitude noise value of 0.5\% (pulse duration noise 0.4\%). NB: The color map has a dynamic range limited to a RIN equal to 0.5\%, meaning RIN data is only visible for wavelengths 800-1430 nm.}
\end{figure}

A very interesting and counter-intuitive fact seen in Figs. 3(a) and 4(a) is that across the bandwidth, away from the edges, the noise level of the generated ANDi SC is lower than the laser amplitude noise imposed on the initial condition. This has in fact also been observed experimentally in a recent work on ANDi SC generation with a 1550 nm mode-locked laser \cite{Shreesha}, but never explained. To explain it we need to look deeper into the SPM-based SC generation and the effects of the anti-correlated amplitude and pulse duration noise separately and together. \\

Let us consider the exact solution for SPM, which with our initial condition is given by $A(z,t)=\sqrt{P_0} {\rm sech}(t/T_0)\exp(i\gamma P_0 {\rm sech}^2(t/T_0) z)$. Expanding the sech$^2$ function in the exponent around its maximum derivative, which corresponds to the maximum frequency shift, one finds $\phi(x)={\rm sech}^2(x)\approx \phi_0+\phi_1(x-x_0)+\phi_2(x-x_0)^2$, where $x_0=0.66$, $\phi_1=-0.77$, and $\phi_2=0$. This gives the Fourier transform
\begin{equation}
    |\tilde{A}(\omega)|^2= P_0T_0^2 \left| \int_{0}^{+\infty} {\rm sech}(x) \left( e^{i(\omega-\omega_0+\Delta \omega)T_0x} + e^{-i(\omega-\omega_0+\Delta \omega)T_0x}  \right)dx \right|^2,
\end{equation}
where the frequency shift is given by $\Delta\omega=-0.77\gamma P_0 z/T_0$. This expression can be used to study the RIN due to amplitude and pulse duration noise, which is basically how much $|\tilde{A}(\omega)|^2$ changes under a change in $P_0$ and/or $T_0$. \\

First of all we see that the frequency shift increases with increasing peak power and/or decreasing pulse duration, as is well-known. Thus an increasing (decreasing) peak power and correlated decreasing (increasing) pulse duration will act together on the frequency shift and lead to a strongly increasing (decreasing) frequency shift. This will in itself lead to a periodic change in the spectral intensity $|\tilde{A}(\omega)|^2$ and thus a periodic RIN. 

\begin{figure}[t]
   \begin{minipage}[c]{.46\linewidth}
   \includegraphics[scale=0.63]{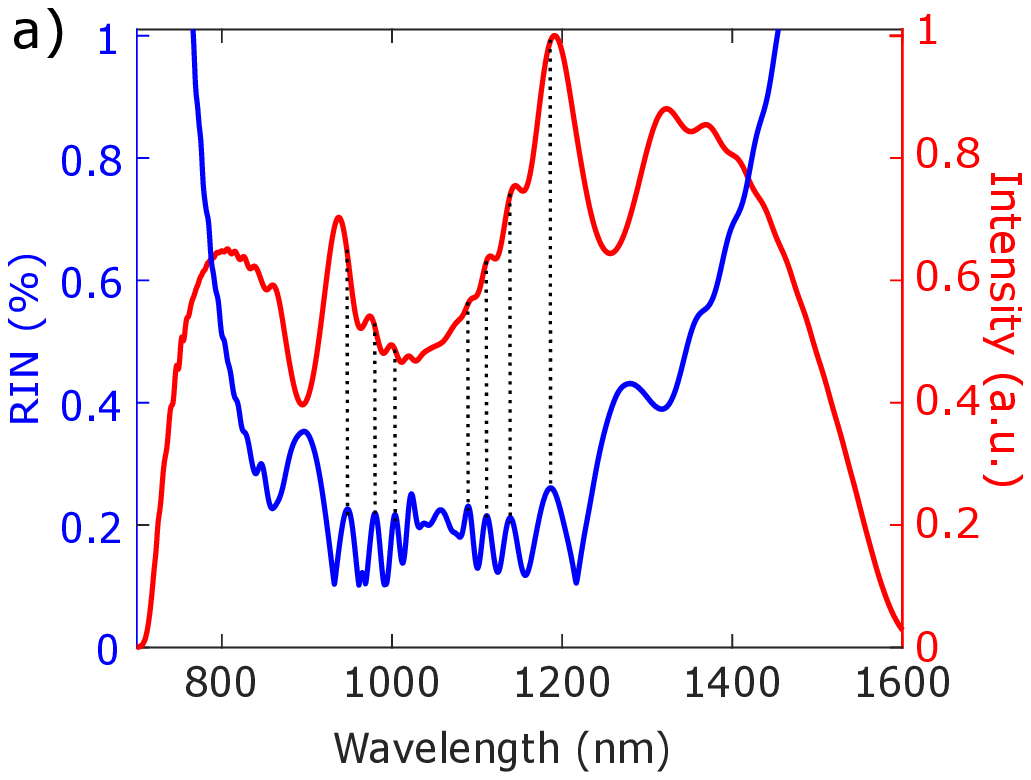}  
   \end{minipage} \hfill
   \begin{minipage}[c]{.48\linewidth}
\includegraphics[scale=0.63]{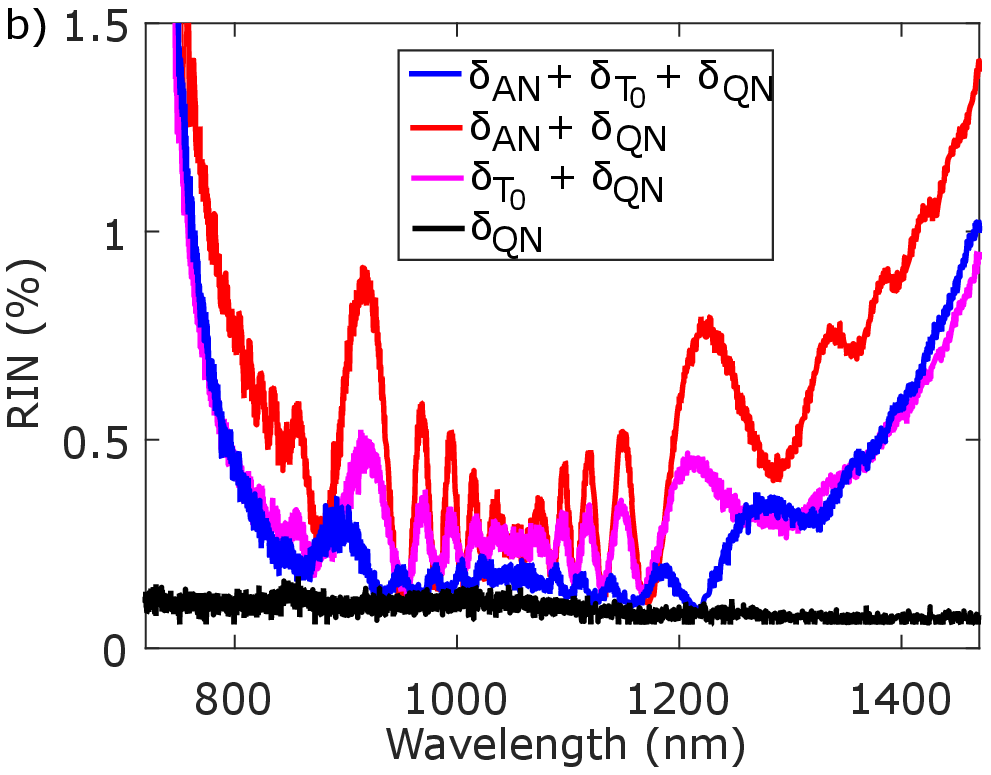}
   \end{minipage} 
\caption*{Fig 4.(a) RIN profile (blue line) and mean spectrum (red line) of an ensemble of 20 pulses after 10 cm of fiber with 100 kW peak power, 50 fs pulse duration at 1054 nm for an amplitude noise value of 0.5~\% (pulse duration noise of 0.4\%). (b) RIN spectrum as a function of the input noise : OPM only (black line), amplitude noise plus OPM noise (red line), pulse duration noise plus OPM (pink line) and amplitude noise plus pulse duration noise and OPM (blue line).}
\end{figure}

Let us now look at two simple special cases: (1) At the input $z=0$ the frequency shift is zero and the solution becomes $|\tilde{A}(\omega)|^2= \pi^2P_0T_0^2\textrm{sech}.^{2}(\pi T_0[\omega-\omega_0]/2)$. Since $T_0$ will decrease when $P_0$ increases (and vice versa) according to Eq. (2), we see that at the center frequency the amplitude and anti-correlated pulse duration noise act against each other, i.e., they will tend to eliminate each other. (2) If we consider sufficiently large propagation distances, so that the frequency shift $\Delta\omega$ is large, then we can neglect the first (second) exponential in Eq. (6) when $\omega\approx\omega_0+\Delta\omega$ ($\omega\approx\omega_0-\Delta\omega$), because the exponential will be rapidly oscillating. In other words, the two outer slopes in the SPM spectrum can be treated in isolation. In this case the approximate solution for the long wavelength SPM slope is $|\tilde{A}(\omega)|^2= (\pi^2/4)P_0T_0^2\textrm{sech}.^{2}(\pi T_0[\omega-\omega_0+\Delta\omega]/2)$. At the peak we therefore see that again the amplitude and anti-correlated pulse duration will act oppositely and tend to cancel each other. The fact that the RIN is lower than the pump laser amplitude noise is thus to be expected from theory and due to the anti-correlated pulse duration noise. \\

Finally, Fig. 4(b) compares the impact of the different noise sources including OPM, pulse duration and amplitude noise, on the RIN spectrum. We clearly see when all the noise sources are included (blue curve) the RIN level is lower than in isolation (red and pink curves), expect for OPM noise (black curve). \\
\section{Conclusion}

We have presented a detailed numerical study of the impact of pump laser amplitude noise on the coherence of the SC generated in ANDi PCFs with femtosecond high peak power mode-locked pump lasers. In particular, we have shown that considering nominal values of amplitude noise drastically affects the SC coherence on the spectral edges. Indeed, when only one-photon-per-mode quantum noise is taken into account, the coherence first starts to degrade for pulse durations above 1.2 ps, while if a weak pump laser amplitude noise of 0.5~\% is taken into account, the degradation starts already at a pulse duration of $\sim$ 50~fs. \\

We have looked into the specific spectral profile of the RIN of a typical low-noise ANDi SC (50 fs pulse duration, 100 kW peak power) and found that it is strongly increasing towards the spectral edges of the SC, as expected, but much less so on the red edge than the blue edge. We found that this is due to the noise suppression effect of the long wavelength confinement loss edge of the ANDi PCF, occuring already at 1450 nm due to the small holes of the typical ANDi PCF we considered.\\

In the central part of the low-noise ANDi SC we demonstrated that the peaks in the SC spectrum are correlated with the peaks in the RIN spectrum and that this correlation is especially apparent at shorter fiber length where SPM is dominating and the RIN profile is periodic. In particular, we demonstrated numerically that the SC noise in the central part is lower than the considered pump laser amplitude noise and that this is due a competition between the amplitude and anti-correlated pulse duration noise, which in combination gives a lower noise than in isolation. We confirmed analytically that this should be so and that it is due to the anti-correlation of the amplitude and pulse duration of the pump laser. \\

Our study of the absolute values and finer details of SC noise in ANDi fibers is of substantial value to potential applications, such as OCT and metrology, which require ultra-low-noise SC light sources. Indeed, this study constitutes the first in-depth look into the effect of technical noise sources on the ANDi SC process and provides grounds for further research to achieve a better understanding of these physically complex processes.

\section*{Acknowledgment}

The authors thanks A. Heidt for his helpful discussions. The research project leading to this work has received funding from the European Union's Horizon 2020 research and innovation programme under the Marie Sklodowska-Curie grant agreement No 722380 (the project SUPUVIR), and grant agreement No 7326 (the project GALAHAD). T. Sylvestre and J. Dudley thank the support of Agence Nationale de la Recherche (ANR-17-EURE-0002, ANR-15-IDEX-0003).

\end{document}